\documentclass[baaa]{baaa}

\usepackage[pdftex]{hyperref}
\usepackage{subfigure}
\usepackage{natbib}
\usepackage{helvet,soul}
\usepackage[font=small]{caption}

\begin{document}


\journalvol{58}
\journalyear{2015}
\journaleditors{P. Benaglia, D.D. Carpintero, R. Gamen \& M. Lares}


\contriblanguage{0}
\hyphenation{es-tre-llas cons-truc-cion co-rres-pon-den des-cri-bi-mos cons-ti-tu-yen-te pers-pec-ti-vas nues-tro cons-tan-te ba-rras}


\contribtype{1}

\thematicarea{7}

\title{Estrellas h\'ibridas:}
\subtitle{una aproximaci\'on semi-anal\'itica a T finita}


\titlerunning{Estrellas h\'ibridas: una aproximaci\'on semi-anal\'itica a T finita}


\author{M. Mariani\inst{1, 2} \& M. Orsaria\inst{1, 2}}
\authorrunning{Mariani \& Orsaria}


\contact{mariani@fcaglp.unlp.edu.ar}

\institute{Grupo de Gravitaci\'on, Astrof\'isica y Cosmolog\'ia, Facultad de Ciencias Astron\'omicas y Geof\'isicas, Universidad Nacional de La Plata \and
  Consejo Nacional de Investigaciones Cient\'ificas y T\'ecnicas (CONICET)
}


\resumen{
A partir de la construcci\'on semi-anal\'itica de una ecuaci\'on de estado (EdE) que tiene en cuenta materia nuclear y de quarks a temperatura finita, estudiamos la posibilidad de que las proto-estrellas de neutrones, sean proto-estrellas h\'ibridas cuyos n\'ucleos est\'an formados por materia de quarks. Obtenemos la relaci\'on masa-radio y discutimos las restricciones recientes de masas y radios para las estrellas de neutrones, considerando los p\'ulsares PSR J1614-2230 y PSR J0348+0432.}

\abstract{Starting from the semi-analytic construction of a equation of state (EoS) which takes into account nuclear and quark matter at finite temperature, we study the possibility that proto-neutron stars, be proto-hybrid stars whose cores are composed of quark matter. We obtain the mass-radius relationship and discuss the latest constraints on masses and radii of neutron stars, considering the pulsars PSR J1614-2230 and PSR J0348 + 0432.
}



\maketitle

\section{Introducci\'on}
\label{S_intro}

Las estrellas de neutrones son objetos compactos que pueden ser utilizados como laboratorios astrof\'isicos ideales para el estudio de temas fundamentales de la astronom\'ia y la f\'isica de part\'iculas. Dichos temas incluyen la f\'isica gravitacional en el regimen de campo fuerte, la influencia de campos magn\'eticos intensos, la superfluidez y la superconductividad, las fuerzas nucleares sometidas a condiciones extremas y las posibles transiciones de fase de la materia densa. 

La reciente detecci\'on de dos estrellas de neutrones masivas, PSR J1614-2230 ($1.97 \pm 0.04 \text{ M}_{\odot}$)\citep{Demorest:2010bx} y PSR J0348+0432 ($2.01 \pm 0.04  \text{ M}_{\odot}$) \citep{Antoniadis:2013pzd} cuyas masas fueron determinadas con gran precisi\'on, permite poner a prueba y acotar los modelos de materia nuclear  para la construcci\'on de la ecuaci\'on de estado (EdE). 

A partir de la determinaci\'on de la masa de estos p\'ulsares, los astrof\'sicos debieron replantearse los modelos te\'ricos para la descripci\'on microsc\'opica de la materia en el interior de las  estrellas de neutrones. Alguna de las alternativas para la construcci\'on de la EdE incluye la posibilidad de que estos objetos contengan hiperones \citep{Yamamoto:2014jga}, bariones formados por tres quarks, o bien que dichos hiperones esten contenidos en sus n\'ucleos \citep{Bednarek:2011gd}. Considerando la presencia de estos bariones, las estrellas de neutrones masivas ser\'ian posibles, 
siempre y cuando hubiese un control de la compresibilidad de la materia a trav\'es del ajuste de las interacciones entre las 
part\'iculas involucradas.

Otra posibilidad es que las estrellas de neutrones contengan materia de quarks, en la que sea viable ajustar alguno de los par\'ametros que media la interacci\'on entre ellos \citep{Bonanno:2011ch,Orsaria:2013hna}. Los modelos m\'as utilizados para el estudio de la materia de quarks en los objetos compactos son el modelo de bolsa del MIT, {\it Massachusetts Institute of Technology}, \citep{Chodos:1974je} y el de Nambu Jona-Lasinio \citep{Nambu:1961tp}. Recientemente, ha comenzado a usarse el M\'etodo del Campo Correlacionador (MCC) que es una aproximaci\'on no perturbativa de la Cromodin\'amica Cu\'antica, que incluye desde los primeros principios la din\'amica del confinamiento\citep{DiGiacomo2002319}.

En este trabajo, construimos la EdE y analizamos la relaci\'on masa-radio (M-R) de estrellas h\'ibridas, objetos compactos formados por un n\'ucleo de quarks rodeado de una corteza hadr\'onica, considerando el MCC para la descripci\'on de la fase de quarks a temperatura finita. En la Secci\'on \ref{EdeH} describimos los modelos de EdE utilizados para la materia de quarks y la materia hadr\'onica. En la Secci\'on \ref{hyb_star} explicamos la construcci\'on del diagrama M-R para las estrellas h\'ibridas y mostramos los resultados. Por \'ultimo, en la Secci\'on \ref{fin} presentamos algunas conclusiones y perspectivas.

\section{EdE h\'ibrida a temperatura finita} 
\label{EdeH}

Para el tratamiento del sistema a temperatura finita, trabajamos con unidades naturales, es decir que $c=\hbar=k_B=1$. De esta manera, la temperatura quedar\'a expresada en dimensiones de energ\'ia, i.e. MeV.

En su etapa inicial, una proto-estrella de neutrones alcanza temperaturas del orden de los $50 \text{ MeV}$ \citep{Lattimer:2004pg} y por un tiempo de lrededor de un minuto permanece caliente y opaca a los neutrinos. Luego, se vuelve transparente  a los neutrinos, transform\'andose en una estrella e neutrones fr\'ia ($T \sim 1 \text{ MeV}$).

En este trabajo, si bien tendremos en cuenta el efecto de la temperatura, no consideraremos la contribuci\'on de los neutrinos en nuestra EdE h\'ibrida. 

\subsection{EdE para los quarks: M\'etodo del Campo Correlacionador}

Para la descripci\'on de la materia de quarks $u$, $d$ y $s$ utilizamos el MCC, recientemente aplicado para el estudio de estrellas h\'ibridas a temperatura cero \citep{Plumari:2013ira, Logoteta:2014xea, Burgio:2015zka}. 

El MCC se parametriza en funci\'on del condensado glu\'onico $G_2$ (que tiene en cuenta efectos no perturbativos de la teor\'ia y caracteriza la fase normal de la materia de quarks) y el potencial quark-antiquark $V_1$ (que tiene en cuenta el confinamiento). Se estima un valor para $G_2 \simeq 0.012 \text{ GeV}^4$, con un $50\%$ de incerteza \citep{Burgio:2015zka}, mientras que $10\text{ MeV}<V_1<100\text{ MeV}$, como en las referencias \cite{Plumari:2013ira, Logoteta:2014xea,  Burgio:2015zka}. 

En el marco de este modelo, la presi\'on del plasma de quarks y gluones resulta:

\begin{equation}
	P_{qg} = \sum_{i=u,d,s} P_i + P_g - \frac{9}{64} G_2 \, , 
\label{pqg}
\end{equation}
donde $P_i$ es la presi\'on de los quarks m\'as la de los antiquarks dada por 

\begin{eqnarray}
\frac{\pi^2}{T^4} P_i  & =  & \frac{\pi^2}{T^4} (P_q + P_{\bar{q}})  \\
 & = & \phi_\nu (\frac{\mu_q - V_1/2}{T}) +
\phi_\nu (\frac{-\mu_q - V_1/2} {T})\,  \nonumber ,	
\end{eqnarray}
con 
\begin{equation}
\phi_\nu (a) = \int_0^\infty du \frac{u^4}{\sqrt{u^2+\nu^2}} 
               \frac{1}{\exp{[\sqrt{u^2 + \nu^2} - a]} + 1}\, , 	
\end{equation}
y $\nu=m_q/T$, donde $\mu_q$ y $m_q$ son el potencial qu\'imico y la masa de los quarks respectivamente, $T$ es la temperatura y $u$ es una variable de integraci\'on. La presi\'on de los gluones es

\begin{equation}
	\frac{\pi^2}{T^4} P_g = \frac{8}{3} \int_0^\infty  d\chi \chi^3 \frac{1}{\exp{(\chi + \frac{9 V_1}{8T} )} - 1}  \, .
\end{equation} 

Aqu\'i, $\chi$ representa una variable de integraci\'on.

Dada la Ec.\ref{pqg}, y siguiendo la referencia \cite{Masperi:2004pb}, realizamos un doble desarrollo en serie de potencias, en t\'erminos de $m_q^2/(u^2T^2+m_q^2)^2$ y de $(\mu_q-V_1/2)/T$. Mediante este m\'etodo, obtuvimos un resultado anal\'itico para calcular las magnitudes termodin\'amicas a temperatura finita.

En el contexto estelar, consideramos equilibrio $beta$ entre las part\'iculas y conservaci\'on de carga el\'ectrica local dada por

\begin{equation}
	2n_u - n_d - n_s - 3n_e = 0 \, ,
\end{equation}
donde $n_{i=u,d,s,e}$ son las densidades num\'ericas de los quarks y los electrones.

A partir de este planteo, es posible obtener la EdE para la materia de quarks y gluones en la estrella h\'ibrida.  

\begin{figure}[!ht]
  \centering
  \includegraphics[width=0.35\textwidth]{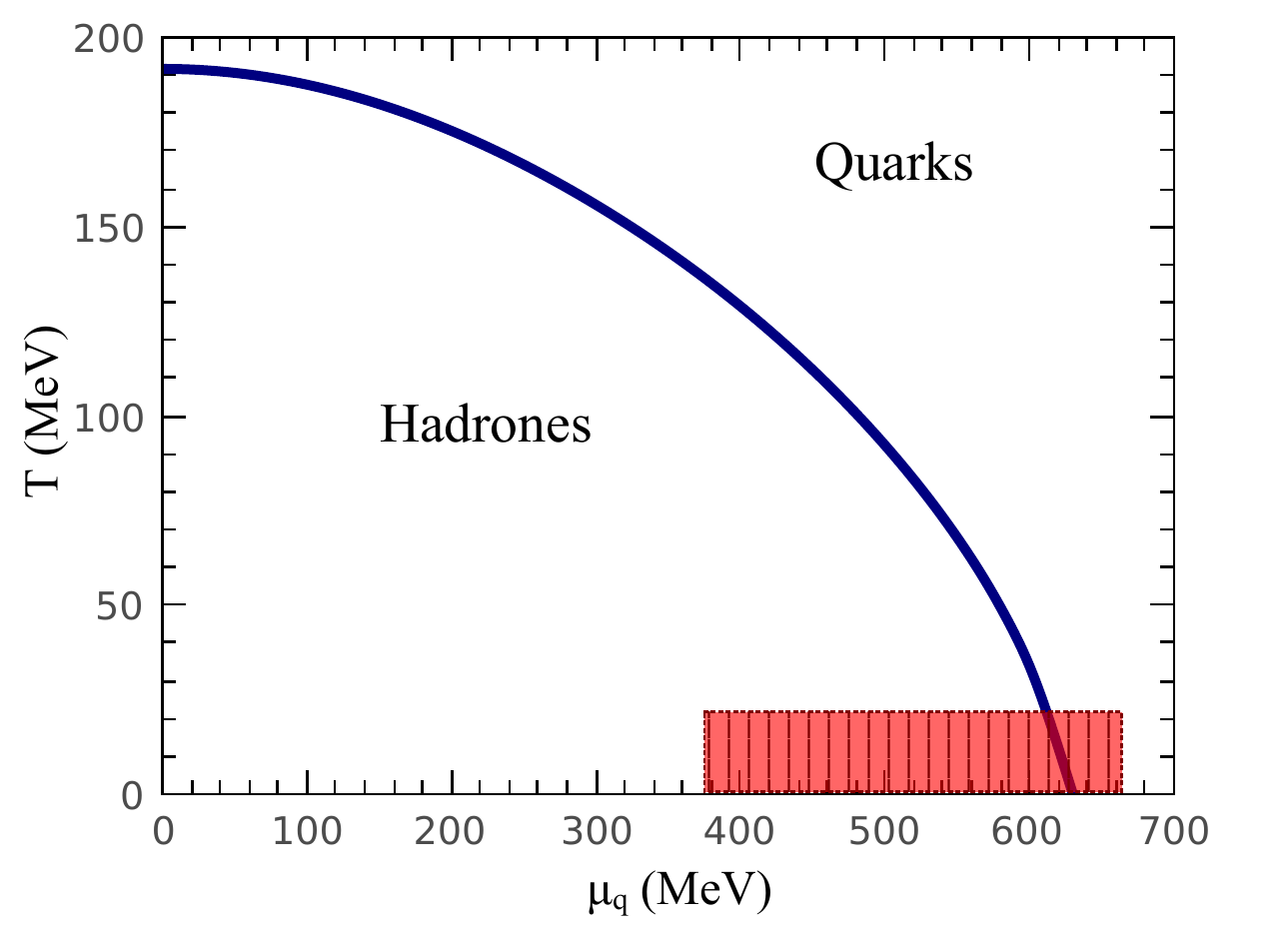}
  \caption{Esquema del diagrama de fases de la Cromodin\'amica Cu\'antica obtenido mediante el MCC. La zona que corresponde a las estrellas de neutrones se indica mediante el recuadro rayado.}
  \label{phase_diag}
\end{figure}

A fin de estudiar el comportamiento del MCC y reproducir las caracter\'isticas principales del diagrama de fases de la Cromodin\'amica Cu\'antica, calculamos la curva de igualdad de presi\'on entre las fases de quarks y  de hadrones en el plano $T-\mu_q$ como muestra la Fig.\ref{phase_diag}. Los resultados resultan compatibles con aquellos expuestos en \cite{Simonov:2007xc}.

\subsection{EdE para los hadrones}

En el caso de la materia hadr\'onica, utilizamos una tabla disponible en la literatura para obtener la EdE a temperatura finita \citep{Shen:2010jd, Shen:2011kr}. En esta tabla se considera la aproximaci\'on de Campo Medio Relativista para modelar la materia formada por neutrones, protones y leptones. Tambi\'en, en este caso se considera equilibrio $beta$ y neutralidad de carga el\'ectrica para el c\'alculo de las magnitudes termodin\'amicas.

\section{C\'alculo de las estrellas h\'ibridas}
\label{hyb_star}

A partir de las EdE para la materia de quarks y hadr\'onica, estudiamos la posibilidad de una transici\'on de fase hadr\'on-quark.

\begin{figure}[!ht]
  \centering
  \includegraphics[width=0.45\textwidth]{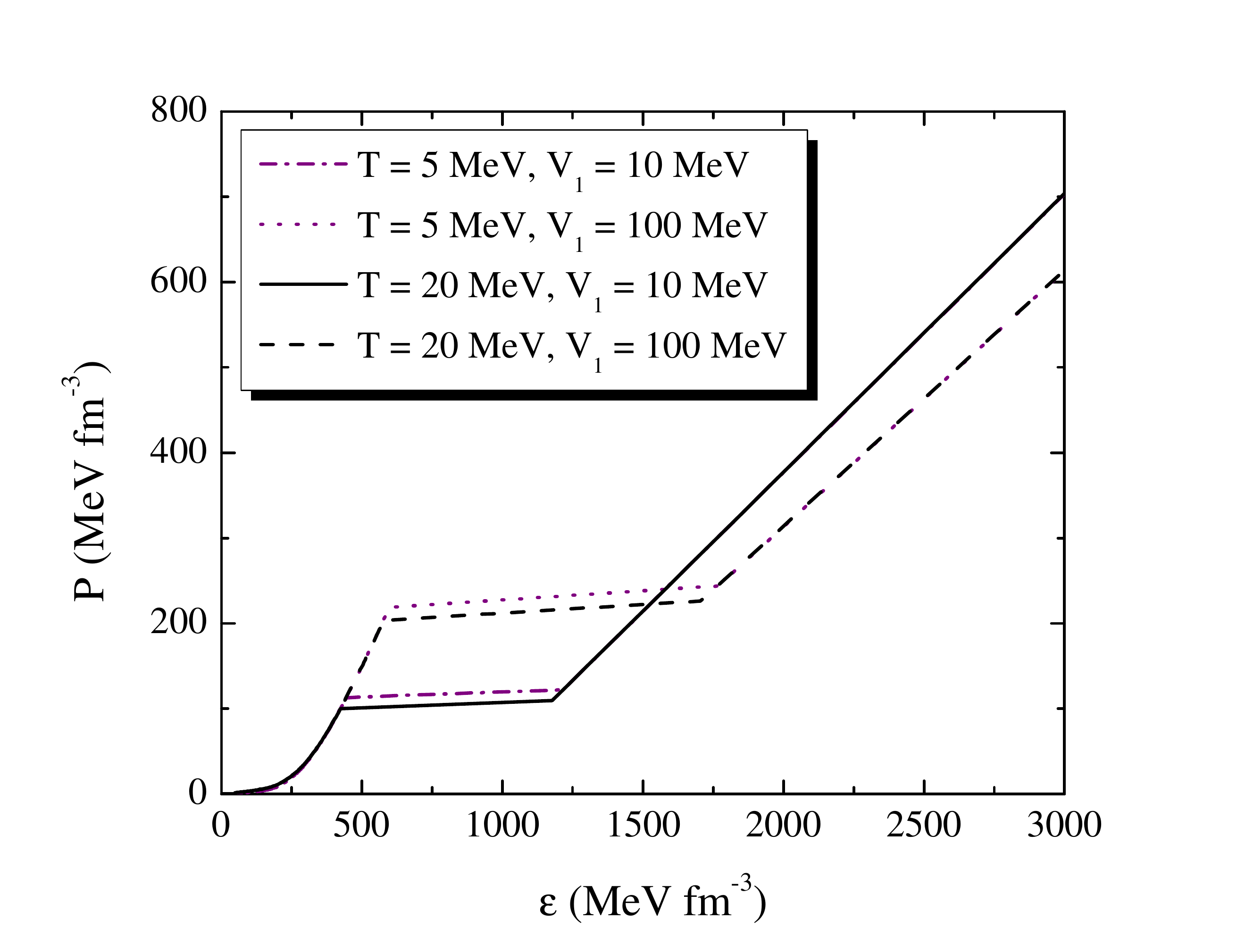}
  \caption{Ecuaciones de estado h\'ibridas para distintos valores de los par\'ametros del MCC. Las regiones de presi\'on constante corresponden a la transici\'on de fase hadr\'on-quark. Los c\'alculos se realizaron con $G_2 = 0.011\text{ GeV}^4$.}
  \label{eos_diag}
\end{figure}

Trabajamos bajo la construcci\'on de Maxwell, en la cual se considera una transici\'on de fase abrupta de primer orden a presi\'on constante y con una carga conservada, $\mu_b$, potencial qu\'imico bari\'onico. De esta manera, obtuvimos EdE h\'ibridas (Fig.\ref{eos_diag}) a partir de las cuales calculamos las configuraciones de equilibrio de la familia de estrellas h\'ibridas. Para ello, integramos simultaneamente la ecuaci\'on de equilibrio hidrost\'atico relativista, ecuaci\'on Tolman-Oppenheimer-Volkoff (TOV), y la ecuaci\'on de continuidad de la masa.

\begin{figure}[!ht]
  \centering
  \includegraphics[width=0.45\textwidth]{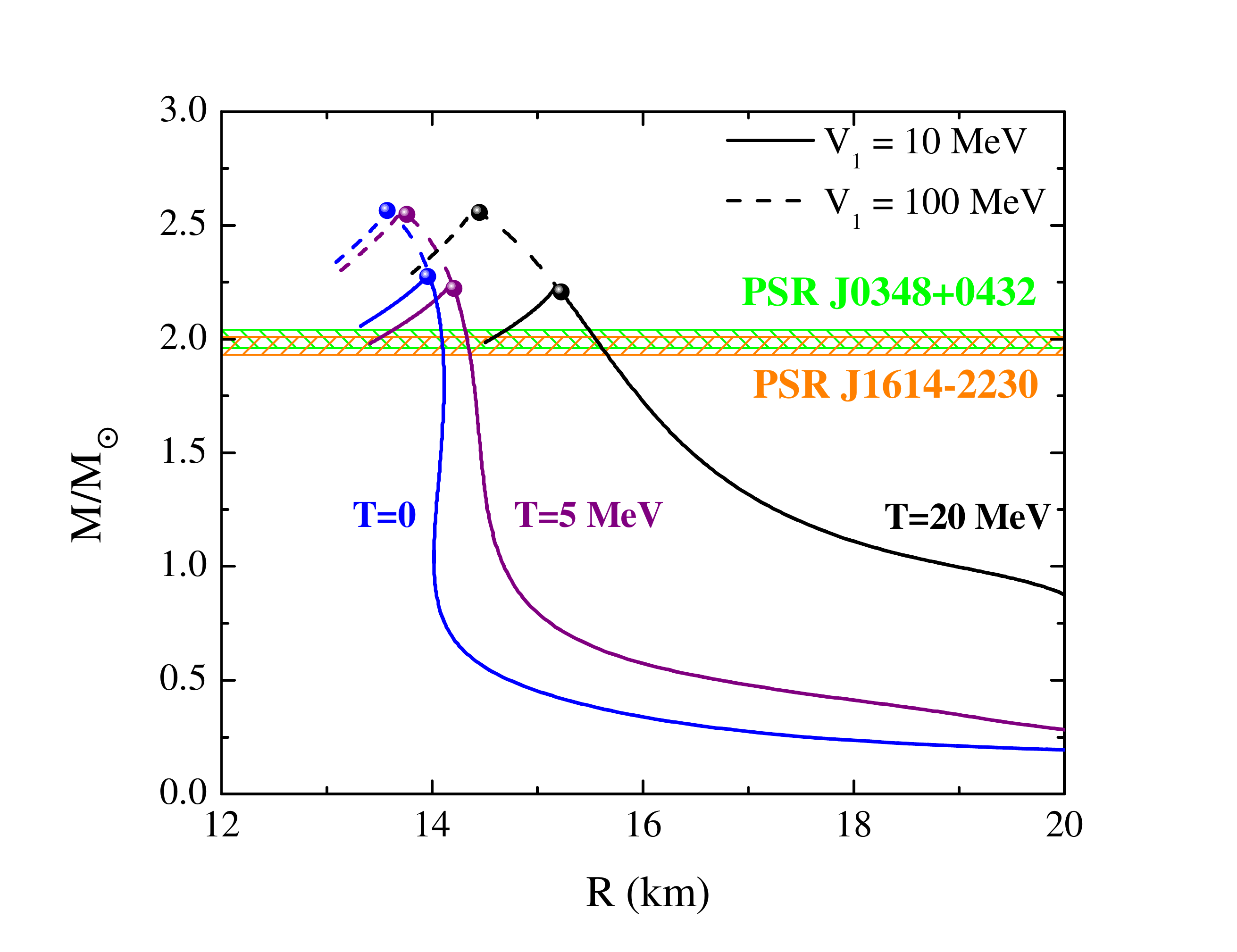}
  \caption{Relaci\'on M-R para estrellas h\'ibridas. Las curvas muestran soluciones de equilibrio hidrost\'atico para dos valores del parametro $V_1$ del MCC y diferentes temperaturas. Luego de alcanzar la masa m\'axima, las soluciones son inestables a medida que el radio disminuye. El punto en cada curva indica la regi\'on a partir de la cual se encuentran las estrellas h\'ibridas. Las curvas correspondientes a $T=0$ permiten realizar la contrastaci\'on con las masas observadas de los p\'ulsares PSR J1614-2230 y PSR J0348+0432, representadas por las barras horizontales.}
  \label{mr_diag}
\end{figure}

Al resolver TOV obtuvimos curvas de soluciones estables en el plano Masa-Radio (M-R) para diferentes valores del conjunto de par\'ametros de nuestro modelo: consideramos estrellas isotermas con temperaturas de $T = 5, 20\text{ MeV}$; tambi\'en, usamos  el parametro $V_1 = 10, 100\text{ MeV} $ y mantuvimos constante el par\'ametro $G_2 = 0.011 GeV^4$. Los resultados para la relaci\'on M-R se muestran en la Fig.\ref{mr_diag}.

\section{Conclusiones y perspectivas}
\label{fin}

La implementaci\'on del desarrollo en serie para soluciones semi-anal\'iticas en el marco del MCC permite calcular las EdE del plasma de quarks-gluones para un rango extendido de temperaturas. Adem\'as, mediante este tratamiento logramos reproducir consistentemente el diagrama de fases simplificado de la Cromodin\'amica Cu\'antica.

Por otro lado, la aplicaci\'on de este m\'etodo al caso de estrellas h\'ibridas permite obtener resultados que reproducen los valores de las masas observados recientemente para estrellas de neutrones. En este sentido, resulta relevante notar que las curvas M-R obtenidas son comparadas con las observaciones de los p\'ulsares PSR J1614-2230 y PSR J0348+0432 (Fig.\ref{mr_diag}). En nuestro caso, los valores seleccionados para los par\'ametros del MCC permiten obtener masas m\'aximas mayores que $2 \text{ M}_{\odot}$.


Respecto de los radios obtenidos para las diferentes familias de estrellas, tuvimos en cuenta la cota inferior de \cite{Chen:2015zpa}, que establece que las estrellas de neutrones de $1.4\text{ M}_{\odot}$ deben tener radios mayores que $R=10.7\text{ km}$ para no violar la causalidad.

Es importante señalar que ser\'a necesario incorporar la presencia de neutrinos como producto de la interacci\'on d\'ebil y considerar 
el caso de estrellas h\'ibridas no isotermas, donde la entrop\'ia por bari\'on ser\'a constante en las diferentes etapas de enfriamiento de la proto-estrella h\'ibrida \citep{Steiner:2000bi}. Una vez realizadas estas modificaciones, se deber\'a recorrer en forma detallada el espacio de los par\'ametros del modelo y analizar los nuevos resultados obtenidos. 


\begin{acknowledgement}
Los autores agradecen al CONICET y a la UNLP por el apoyo financiero. Tambi\'en agradecen la minuciosa revisi\'on del manuscrito por parte del \'arbitro.
\end{acknowledgement}


\bibliographystyle{baaa}
\small
\bibliography{biblio_mariani}
 
\end{document}